\documentclass[aps,prfluids,onecolumn,amsmath,amssymb,superscriptaddress]{revtex4-2}
\pdfoutput=1

\usepackage{graphicx}				
\usepackage[hidelinks]{hyperref}	
	\hypersetup{pdfauthor={Thomas C. Sykes et al.},pdftitle={Surface jets and internal mixing during the coalescence of impacting and sessile droplets}}
\usepackage{nicefrac}				
\usepackage[separate-uncertainty,multi-part-units=single]{siunitx}	
\usepackage{xcolor}

\newlength{\colwidth}
\newcommand{\panela}{(a)}
\newcommand{\panelb}{(b)}
\newcommand{\panelc}{(c)}
\newcommand{\paneld}{(d)}
\newcommand{\panele}{(e)}

\DeclareSIUnit\px{pixels}			

\begin{document}

\title{Surface jets and internal mixing during the coalescence of impacting and sessile droplets}


\author{Thomas C. Sykes}
\email{mm13tcs@leeds.ac.uk}
\affiliation{EPSRC Centre for Doctoral Training in Fluid Dynamics, University of Leeds, Leeds LS2 9JT, United Kingdom}

\author{Alfonso A. \surname{Castrej\'on-Pita}}
\affiliation{Department of Engineering Science, University of Oxford, Oxford OX1 3PJ, United Kingdom}

\author{J. Rafael \surname{Castrej\'on-Pita}}
\affiliation{School of Engineering and Materials Science, Queen Mary, University of London, London E1 4NS, United Kingdom}

\author{David~Harbottle}
\affiliation{School of Chemical and Process Engineering, University of Leeds, Leeds LS2 9JT, United Kingdom}

\author{Zinedine Khatir}
\affiliation{School of Mechanical Engineering, University of Leeds, Leeds LS2 9JT, United Kingdom}

\author{Harvey M. Thompson}
\affiliation{School of Mechanical Engineering, University of Leeds, Leeds LS2 9JT, United Kingdom}

\author{Mark C. T. Wilson}
\email{M.Wilson@leeds.ac.uk}
\affiliation{School of Mechanical Engineering, University of Leeds, Leeds LS2 9JT, United Kingdom}

\date{\today}

\begin{abstract}
	The internal dynamics during the coalescence of a sessile droplet and a subsequently deposited impacting droplet, with either identical or distinct surface tension, is studied experimentally in the regime where surface tension is dominant. Two color high-speed cameras are used to capture the rapid internal flows and associated mixing from both side and bottom views simultaneously by adding an inert dye to the impacting droplet. Given sufficient lateral separation between droplets of identical surface tension, a robust surface jet is identified on top of the coalesced droplet. Image processing shows this jet is the result of a surface flow caused by the impact inertia and an immobile contact line. By introducing surface tension differences between the coalescing droplets, the surface jet can be either enhanced or suppressed via a Marangoni flow. The influence of the initial droplet configuration and relative surface tension on the long-term dynamics and mixing efficiency, plus the implications for emerging applications such as reactive inkjet printing, are also considered.
\end{abstract}


\maketitle



\section{Introduction}

Droplet coalescence is a pivotal feature in many natural and applied phenomena, including raindrop formation in clouds, inkjet printing and phase-change heat transfer technologies \cite{Cha2016,Khatir2016}. Within the past half-century, the external dynamics of droplet coalescence have been studied extensively, from the growth of a meniscus bridge between coalescing droplets \cite{Wu2004} to pinch-off and satellite formation \cite{Zhang2009}, which may repeat numerous times to form a coalescence cascade \cite{Harbottle2011,Thoroddsen2000}. Nevertheless, both the conditions required for the coalescence of colliding droplets \cite{Al-Dirawi2019} and the physical mechanism initiating coalescence \cite{Perumanath2019} are current areas of research.

Effective mixing between miscible fluids contained within each coalescing droplet (known as the precursor droplets) is required in many applications, such as biochemical reagents in lab-on-a-chip microfluidic devices and chemical reactants in advanced manufacturing technologies like reactive inkjet printing \cite{Wilson2018}. In some situations, rapid mixing can be achieved by supplying external energy to the coalesced droplet, such as through actuation by electrowetting \cite{Paik2003electrowetting}. These techniques are referred to as active mixing and stretch the internal fluid interface to improve the efficiency of molecular diffusion to homogenize the coalesced droplet. However, the provision of external energy is not always practical, especially in scenarios involving successive coalescence events on a substrate with evolving topology. Therefore the internal flows initiated by coalescence are often solely responsible for determining the distribution of fluid from each precursor droplet (passive mixing). Turbulent internal flow can improve mixing, but is difficult to generate and sustain at typical droplet length scales. Laminar internal flows can include complex flow structures, such as internal jets, that are crucial for enabling efficient mixing within passively mixed systems \cite{Anilkumar1991,Tang2016,Xia2017}.

Coalescence may be initiated during the impact of a falling droplet with a sessile droplet on a substrate, which is the typical configuration in inkjet printing. For millimetric droplets with identical fluid properties, similar volumes and inertial dimensionless numbers matched to typical inkjet values, experiments have demonstrated no discernible mixing within the coalesced droplet \cite{Castrejon-Pita2013}. This conclusion is robust to lateral separation between the precursor droplets and has been corroborated by numerical simulations for substrates of various wettabilities \cite{Raman2016successive,Raman2018}. Improved advective mixing can be achieved by the formation of a vortex ring if the sessile droplet is much larger than the impacting droplet. Vortex rings can be formed in a similar manner during the impact of a droplet onto a deep pool \cite{Saha2019}, whereas capillary wave dynamics influence mixing considerably for shallow pools \cite{Ersoy2019}. However, droplet-pool coalescence is critically different from the coalescence of droplets on a substrate due to the absence of a contact line.

An intrinsic feature of many applications is that the precursor droplets consist of different fluids, where differences in the fluid properties can influence the internal dynamics. For precursor droplets of different densities, a stratified coalesced droplet may be formed by an internal gravity current on a longer time scale than the surface tension induced flow \cite{Zhang2015}. Alternatively, the use of non-Newtonian fluids can lead to intricate internal flow structures and good advective mixing \cite{Sun2015static}. Differences in the rheological properties of Newtonian droplets can be used to control the final internal structure of the coalesced droplet, with the viscosity ratio between an oil droplet and an (immiscible) sessile water droplet defining the maximum penetration depth \cite{Chen2017}. In the context of reactive inkjet printing, some studies have considered mixing between impacting and coalescing micrometric droplets of different reactive fluids, but did not resolve the internal dynamics which would be difficult at this length scale (e.g. Ref.~\cite{Fathi2013}).

Surface tension differences between precursor droplets are particularly significant for the internal dynamics, since surface tension influences the Laplace pressure and surface tension gradients drive Marangoni flow tangential to the interface. Marangoni flow is directed towards regions of high surface tension, so acts to reduce overall surface energy. Both experimental and numerical studies have demonstrated that surface tension differences have a greater influence on advective mixing than geometric differences between the precursor droplets which is mainly a result of Marangoni flow \cite{Blanchette2010,Nowak2017}. Due to interfacial flow, the lower surface tension droplet tends to envelop the higher surface tension droplet after coalescence which can generate an internal jet \cite{Luo2019}. The tangential flow velocity increases linearly for moderate surface tension differences, becoming sublinear for larger differences. The velocity reduces with increasing Ohnesorge number, which is the ratio of viscous to inertial and surface tension forces, as viscous forces retard the motion \cite{Liu2013}. Hence, relatively small surface tension differences may lead to significant changes in the dynamics. Such surface tension differences are usually established using different simple fluids, but they can also be due to surfactants. For surfactants, the solutal Marangoni flow induced may depend on the precise chemical nature of the surfactant which can influence the internal dynamics \cite{Nash2018}. Surface tension differences due to surfactants have been shown to reduce color blur and bleeding in inkjet printed droplets at the boundary between colors of different intensity \cite{Oyanagi2003}.

Many studies involving surface tension differences, including those discussed above, concern droplets within an immiscible, high viscosity outer fluid (typically an oil). In particular, these include droplets confined within a microfluidic channel (confined microfluidics) where the high viscosity of the outer fluid suppresses free surface oscillations through viscous dissipation, reduces the rate of meniscus bridge growth and impedes interfacial flow. In these scenarios, the curvature of the precursor droplets and individual Laplace pressures persist for longer, which promotes internal jet formation, whilst surface flows are diminished. Moreover, the jet morphology and dynamics have been shown to depend on the viscosity ratio between the droplets and outer fluid \cite{Nowak2016}. In cases where the outer fluid flows within the microchannel, the precursor droplet order can affect the internal and interfacial flow \cite{Kovalchuk2019}.

In contrast to confined microfluidics, other microfluidic devices rely on manipulating droplets on a solid substrate, known as open-surface microfludics \cite{Jiao2019}. For these systems, coalescence in a low viscosity gaseous outer fluid (typically air) is of interest. For droplets on a substrate, the contact line dynamics also affect the internal and external dynamics \cite{Lai2010}, where improved advective mixing due to Marangoni flow \cite{Ng2017} and delayed coalescence \cite{Bruning2018} may arise. The initial droplet configuration can influence the dynamics in this case and jet-like internal flows can be generated by recirculation for precursor droplets of either identical or different surface tension \cite{Yeh2013}. With the presence of a free surface open to air, purely interfacial phenomena can arise, such as Marangoni-induced spreading of a droplet impacting a deep pool \cite{Taherian2016}. Both experimental and numerical studies have shown that these impacts can lead to Marangoni-induced droplet ejection \cite{Blanchette2009gradients,Shim2017,Sun2018partial}. For precursor droplets of fluids which undergo a precipitating chemical reaction upon mixing, the magnitude of the surface tension difference can determine the extent of spreading and mixing and hence the precipitate pattern \cite{Jehannin2015}. Complex interfacial flow structures and instabilities may also be generated, such as by evaporation-augmented Marangoni flow during the impact of an alcohol droplet with an (immiscible) oil pool \cite{Keiser2017}. These observations indicate the possible rich internal and interfacial dynamics which could be expected during the coalescence of impacting and sessile droplets of different surface tension.

In this work, the internal and interfacial dynamics (at the free surface) during the coalescence of an impacting droplet with a miscible sessile droplet on a solid, flat substrate is studied by means of color high-speed imaging. Ethanol-water mixtures, with a low proportion of ethanol, were used to ensure the flow was dominated by surface tension and that the surface tension of each precursor droplet could be independently modified, enabling the unexplored influence of surface tension differences to be studied in this experimental configuration. Surfactants were avoided due to the unclear influence of their chemical composition on the dynamics \cite{Nash2018}. By coloring the impacting droplet with an inert dye, the internal dynamics were passively monitored. The use of two high-speed cameras to acquire two perspectives (side and bottom) simultaneously allowed internal and interfacial phenomena to be distinguished, enabling an accurate assessment of advective mixing to be made. The influence of lateral separation and surface tension differences is considered to elucidate both the initial internal and interfacial dynamics, in addition to the longer-term mixing efficiency.

\section{Experimental details}

\subsection{Materials and characterization}
\label{sec:materials}

Fluid mixtures were prepared from ethanol ($\geq 99.8\%$ purity, Sigma-Aldrich) and deionized water, with the fluid properties given in Table~\ref{tab:mixtureProperties}. All mixture proportions are specified by mass. The surface tension of each mixture was measured using a pendant droplet tensiometer (Biolin Scientific Theta T200) by forming the largest sustainable droplet (\SIrange{7}{13}{\micro\litre}) at the end of a stainless steel blunt end dispensing tip (Fisnar 22 gauge), within a sealed environment. The pendant droplet was analyzed for \SI{60}{\second} in each measurement (repeated at least four times), with its volume being automatically maintained by infusing additional fluid through the dispensing tip. Additionally, surface tension was verified using a bubble pressure tensiometer (SITA pro line t15). The surface tension measured was consistent with Ref.~\cite{Vazquez1995}. The error reported combines the random measurement error ($\pm \SI{0.2}{\milli\newton\per\metre}$) and the random error due to variations in each sample. To visualize the internal flow, a small amount (approximately 100~ppm) of Malachite green dye (Sigma-Aldrich) was added to the impacting droplet. The amount of dye used was minimized to avoid appreciable changes in the fluid properties, especially surface tension which changed by less than 1\% and within experimental error of the reported values. The density of each mixture was measured using a calibrated \SI{25}{\milli\litre} density bottle with an analytical balance, whereas the viscosity was derived from Ref.~\cite{Khattab2012}.

Visual accessibility from below was achieved by coalescing droplets on glass slides (Fisherbrand plain glass, thickness \SIrange{1}{1.2}{\milli\meter}) which were silanized to increase their hydrophobicity \cite{Arkles2009}. Each substrate consisted of a new glass slide rinsed with Milli-Q water (type 1 ultrapure water) and dried with nitrogen before being placed in a sealed container with \SI{0.5}{\milli\litre} of a silane (dichloromethyl-n-octylsilane, 98\%, Alfa Aesar) to allow vapor deposition for \SIrange{6}{8}{\minute}. The slide was subsequently rinsed and dried prior to use. The equilibrium contact angles of the fluid mixtures on these substrates are reported in Table~\ref{tab:mixtureProperties}. The contact angle measurements were made on a droplet deposited from a dispensing tip consistent with the deposition of the sessile droplet in the coalescence experiments and the contact angle determined by fitting the Young-Laplace equation. The equilibrium contact angle of a water droplet was measured on each substrate produced, with an individual substrate retained only if consistent with Table~\ref{tab:mixtureProperties}. The smallest advancing contact angle, $\theta_a$ was determined by inflating a sessile droplet with additional fluid through an embedded dispensing tip and determining the smallest contact angle for which the contact line moves. Similarly, the largest receding contact angle, $\theta_r$ was determined by deflating a droplet. The measured advancing and receding contact angles were typically $\theta_a \approx \ang{110}$ and $\theta_r \approx \ang{70}$, respectively, so the substrate has a high contact angle hysteresis of approximately $\ang{40}$. During the coalescence events, the contact line generally remains pinned after the initial spreading, and only recedes for very small contact angles. Hence, the substrate can be characterized as strongly pinning (see Ref.~\cite{Bormashenko2013paper}).

\begin{table}[b]
	\caption{Fluid properties of each droplet, with the ensuing experimental conditions. The viscosities were derived from Ref.~\cite{Khattab2012}.\label{tab:mixtureProperties}}
	\begin{ruledtabular}
		\begin{tabular}{lcccc}
			Fluid No.													& 1					& 2				  	& 3 				& 4					\\
			Ethanol Mass \%												& 0.0				& 4.0             	& 8.0             	& 18.0				\\
			Density, $\rho$ (\si{\kilogram\per\cubic\meter})          	& $997 \pm 1$		& $990 \pm 1$     	& $984 \pm 1$     	& $968 \pm 1$		\\
			Viscosity, $\mu$ (\si{\milli\pascal\second})              	& $0.93 \pm 0.01$	& $1.07 \pm 0.03$ 	& $1.20 \pm 0.02$ 	& $1.56 \pm 0.03$	\\
			Surface Tension, $\sigma$ (\si{\milli\newton\per\meter})  	& $72.4 \pm 0.2$	& $58.0 \pm 0.5$	& $50.5 \pm 0.4$	& $39.9 \pm 0.3$    \\
			Equilibrium Contact Angle (degrees)							& $91 \pm 2$		& $82 \pm 2$      	& $74 \pm 2$		& $66 \pm 2$		\\
			Impacting Droplet Radius, $r$ (\si{\milli\meter}) 			& $1.16 \pm 0.02$	& $1.07 \pm 0.02$	& $1.02 \pm 0.02$	& $0.96 \pm 0.02$	\\
			Impacting Droplet Velocity, $u$ (\si{\meter\per\second}) 	& $0.50 \pm 0.04$	& $0.51 \pm 0.06$ 	& $0.50 \pm 0.04$ 	& $0.51 \pm 0.04$
		\end{tabular}
	\end{ruledtabular}
\end{table}

\subsection{Procedure}
\label{sec:procedure}

Each precursor (impacting or sessile) droplet was generated by dripping from a stainless steel blunt end dispensing tip (Fisnar 30 gauge) using a manually controlled syringe pump (World Precision Instruments Aladdin), set at a flow rate of \SI{30}{\micro\litre\per\minute} until the pendant droplet detached due to gravity and fell vertically towards the substrate. Independent, identical dispensing systems (syringe pumps and dispensing tips) were used to generate the undyed sessile and dyed impacting droplets, with the dispensing tips \SI{4}{\milli\metre} apart. The dispensing tip used to generate the sessile droplet was mounted with the blunt end \SI{5.5(5)}{\milli\meter} above the substrate so the droplet was deposited gently and acquired an approximately circular footprint. The dispensing tip used to generate the impacting droplet was mounted higher to achieve a greater impact velocity, with the blunt end \SI{16.5(5)}{\milli\meter} above the substrate. The impacting droplet was always in the deposition regime where it simply spread radially outwards after striking the substrate without any breakup or splashing which would occur for higher impact velocities \cite{Rioboo2001}, as studied by other authors (e.g. Ref.~\cite{Wang2018}). To remove any effect of evaporation at the meniscus of the dispensing tips, an extra droplet was generated (and caught before hitting the substrate) immediately before each precursor droplet was deposited.

\begin{figure}
	\includegraphics[width=1.5\colwidth]{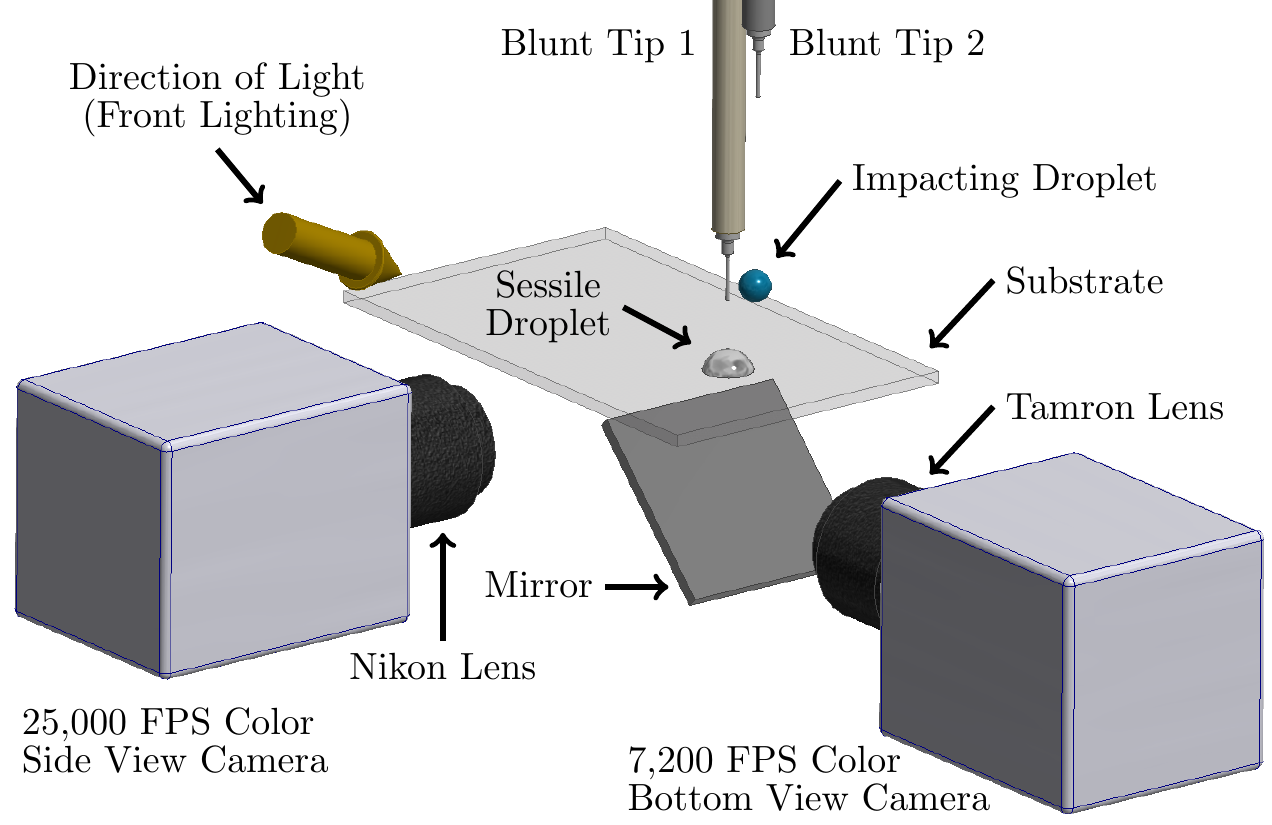}
	\caption{Schematic diagram of the experimental setup. The undyed, sessile droplet is deposited from blunt tip 1; the dyed, impacting droplet is deposited from blunt tip 2. The droplets were front-lit by a constant light source.\label{fig:experimentalSetup}}
\end{figure}

The velocity and radius of the impacting droplet were determined by image processing and are recorded in Table~\ref{tab:mixtureProperties}. These values correspond to the equivalent spherical radius of the precursor sessile droplet (i.e. immediately before it was deposited on the substrate). The deposition of the impacting droplet is dynamically characterized by the Weber, $\textrm{We} = \rho u^2 r/\sigma$ and Ohnesorge numbers, $\textrm{Oh} = \mu/\sqrt{\rho\sigma r}$, where $\rho$, $\sigma$ and $r$ are the density, surface tension and radius of the droplet, respectively. The velocity, $u$ is that of the impacting droplet immediately before landing. In this work, $\textrm{We} \approx 5$ and $\textrm{Oh} \approx \num{5e-3}$ for a typical droplet (i.e. $\rho = \SI{e3}{\kilo\gram\per\meter\cubed}$; $\mu = \SI{e-3}{\pascal\second}$; $\sigma = \SI{50e-3}{\newton\per\meter}$; $r = \SI{e-3}{\meter}$; $u = \SI{0.5}{\meter\per\second}$), which indicates the flow is dominated by surface tension. The equivalent Reynolds number is $\textrm{Re} = \sqrt{\textrm{We}}/\textrm{Oh} \approx 500$. Furthermore, the Bond number is $\textrm{Bo} = gr^2\Delta\rho/\sigma \approx 0.2$, where $g$ is Earth's gravitational acceleration and $\Delta\rho \approx \SI{e3}{\kilo\gram\per\meter\cubed}$ is the density difference between the droplet and surrounding air. The dimensionless numbers indicate that surface tension dominates over gravitational forces despite the relatively large droplet size.

The experimental setup is illustrated in Fig.~\ref{fig:experimentalSetup}. The silanized substrate was mounted as a rigid cantilever on a translation stage providing 2-axis horizontal motion (Comar Optics), with \SI{10}{\micro\meter} precision in each direction. The combined structure was mounted on an elevation stage (Comar Optics), thereby providing the substrate with 3-axis motion. The substrate, supporting the sessile droplet, was conveyed by the translation stage to achieve the desired lateral separation with respect to the subsequently deposited impacting droplet. Droplet positions were determined by two cameras using a long exposure (low light mode) and fiducial markers; a side view gave the lateral separation and a bottom view ensured centerline alignment. The precursor sessile droplet was deposited on the substrate some time prior to coalescence and the volatility of ethanol is higher than water. Experiments were therefore executed expeditiously, with the time between successive droplet depositions kept approximately constant (\SI{20(4)}{\second}) for consistency. Evaporation was quantified by recording the volume loss from a sessile droplet for each fluid mixture with the tensiometer for \SI{50}{\second}. The results show that the volume loss over the period of interest (up to \SI{24}{\second}) is not sufficient to appreciably change the surface tension and therefore does not affect the trends identified in this work (see Supplemental Material for further analysis \cite{SupplMat}). Each fluid mixture was produced on the day of use and the surface tension of a sample was verified using the bubble pressure tensiometer. Each experiment was repeated at least three times to establish the typical dynamics. Coalescence took place in air at room temperature (\SI{23(1)}{\celsius}) and atmospheric pressure.

\subsection{Imaging}

Previous work imaging internal dynamics during droplet coalescence on a substrate has generally been limited to a single perspective, usually with a top or bottom view (e.g. Ref.~\cite{Yeh2015}), but occasionally complemented by a side view (e.g. Ref.~\cite{Castrejon-Pita2013}) or two views for slower dynamics (e.g. Ref.~\cite{Zhang2015}). However, simultaneous imaging has already been shown to be essential for accurately evaluating the extent of mixing within coalesced droplets, for which relatively low frame rates are sufficient \cite{Paik2003electrowetting}. Using two color high-speed cameras to capture both side and bottom views simultaneously, a more complete understanding of the internal dynamics is derived. Moreover, surface and internal dynamics can be distinguished.

In this work, a high-speed camera (a color Phantom v2512) captured the dynamics from the side, using a Nikon AF Micro \SI{60}{\milli\meter} lens with aperture set to f/4. The effective magnification of the lens was increased using extension tubes (Kenko \SI{32}{\milli\meter} and a Nikon K extension ring set) to give a working distance of \SI{37}{\milli\meter}. The pixel resolution was $1024\times768$, yielding an effective resolution of \SI{91.5(5)}{\px\per\milli\meter}. Images were recorded at 25,000 frames per second (FPS), with an exposure of \SI{12}{\micro\second}. To reduce glare around the free surface in this view, the camera was inclined slightly relative to the substrate (approximately \ang{3}).

A second high-speed camera (a color Phantom Miro LAB 310) captured the dynamics from below, through the substrate via an optical mirror (Thorlabs ME2S-G01) mounted \ang{45} to the substrate. This configuration is preferable to a top view, since it clearly captures the droplet footprint on the substrate and avoids distortion from the curved free surface. A fixed aperture macro lens (Tamron SP AF \SI{90}{\milli\metre} f/2.8) was used with two extension tubes (Kenko \SI{20}{\milli\meter} and \SI{12}{\milli\meter}). The pixel resolution was $768 \times 576$, yielding an effective resolution of \SI{65.0(5)}{\px\per\milli\meter}. Images were recorded at 7,200 FPS, with an exposure of \SI{120}{\micro\second}.

The camera arrangement is shown in Fig.~\ref{fig:experimentalSetup}. The cameras were manually triggered by a single \SI{500}{\micro\second} pulse provided directly to each camera by a pulse generator (TTi TGP110). Both cameras were focused on the droplet impact point on the substrate and positioned to fully capture coalescence for all lateral separations studied. A traditional shadowgraph technique is not suitable for the acquisition of color images, so a front-lighting arrangement was used. A single constant light source (89 North PhotoFluor II) was positioned approximately \SI{50}{\milli\meter} from the impact point, to the right of the side view camera's lens and oblique to the horizontal. A white background in each camera view maximized the amount of light reaching the sensors. The light source shutter was only opened for a short and consistent time encompassing coalescence (usually less than \SI{5}{\second}) to maintain a constant temperature environment.

\section{Image processing}
\label{sec:imageProc}

Image processing to track internal and external edges was performed using a custom \textsc{MATLAB} code. Edge detection was preferred to image segmentation (e.g. thresholding) due to apparent color variations within the droplet caused by front-lighting. First, an approximation to the background was subtracted from each frame and the image contrast changed to saturate 1\% of pixels. A Gaussian low-pass filter (standard deviation 2) was then applied via the frequency domain to reduce random noise. Edge pixels were detected using a subpixel edge detection method as suggested by Ref.~\cite{Trujillo-Pino2013}, which is apt for imperfect (realistic) images that may be noisy and have close contours. The detected edge pixels were filtered by the direction of the intensity normal vector and associated with each other based on proximity to determine individual edges. The appropriate internal and external edges were then identified from the set of all edges. The color images acquired allowed the exploitation of the constituent RGB color channels, with the red channel used to distinguish between dyed and undyed fluid (for internal edges), whilst the blue channel enabled each droplet to be identified from the background (for external edges).

The internal fluid interface between the dyed and undyed fluids was exclusively tracked using the bottom view from which a time series of horizontal position (in the plane of the side view) is obtained. For each horizontal position detected, the height of the free surface above the substrate at that location is extracted from the corresponding side view frame. This analysis yields the two-dimensional position of surface phenomena in the plane of the side view. Horizontal positions are matched between the side and bottom views based on the right contact point of the undisturbed sessile droplet. The matched position is confirmed with a fiducial marker on the substrate, from which distances are derived accounting for the different effective resolution of each view. Summarizing, the horizontal position of the internal leading edges were tracked from the bottom view, whilst the corresponding free surface height was acquired from the side view.

The timing is based on the side view (highest frame rate) with each bottom view frame matched to side view times. Due to the high frame rates of both views compared to the time scales of the phenomena studied, the error resulting from the temporal discrepancy is negligible. Time zero is taken as the frame immediately before the first visible contact between droplets. Timing was synchronized by identifying time zero independently in each view.

\section{Droplets with Equal Fluid Properties}

\subsection{Lateral separation}
\label{sec:lateralSeparation}

The impact of a dyed droplet of fluid 2 (see Table~\ref{tab:mixtureProperties}) onto an undyed sessile droplet of the same fluid is shown in Fig.~\ref{fig:lateralSeparation} and the accompanying videos (provided in the Supplemental Material \cite{SupplMat}) for three different lateral separations. For the two smallest lateral separations (Figs.~\ref{fig:lateralSeparation}a and \ref{fig:lateralSeparation}b), the impacting droplet collides with the sessile droplet before the substrate. The requirement for coalescence during this interaction is that the air layer between the droplets drains enough that intermolecular (van der Waals) forces can cause the remaining film of air to rupture. If the air layer does not drain sufficiently during the interaction, then the droplets may bounce without coalescing \cite{Al-Dirawi2019}. Due to the Weber number, there is a small delay (approximately \SI{2}{\milli\second}) between collision and coalescence whilst the entrapped air layer drains at these lateral separations. During this time, the droplet free surface deforms and when coalescence eventually occurs, air is entrained around the internal interface (visible as small bubbles at \SI{4}{\milli\second} in Figs.~\ref{fig:lateralSeparation}a and \ref{fig:lateralSeparation}b). This phenomenon does not influence the long-term internal dynamics and mixing behavior studied here. Air entertainment does not occur when the impacting droplet strikes the substrate first and coalescence is initiated as the impacting droplet spreads across the substrate (e.g. Fig.~\ref{fig:lateralSeparation}c). However, for the axisymmetric case, significant droplet deformation was observed before coalescence occurred at a time critically dependent on the initial conditions.

\begin{figure}
	\includegraphics[width=2.0\colwidth]{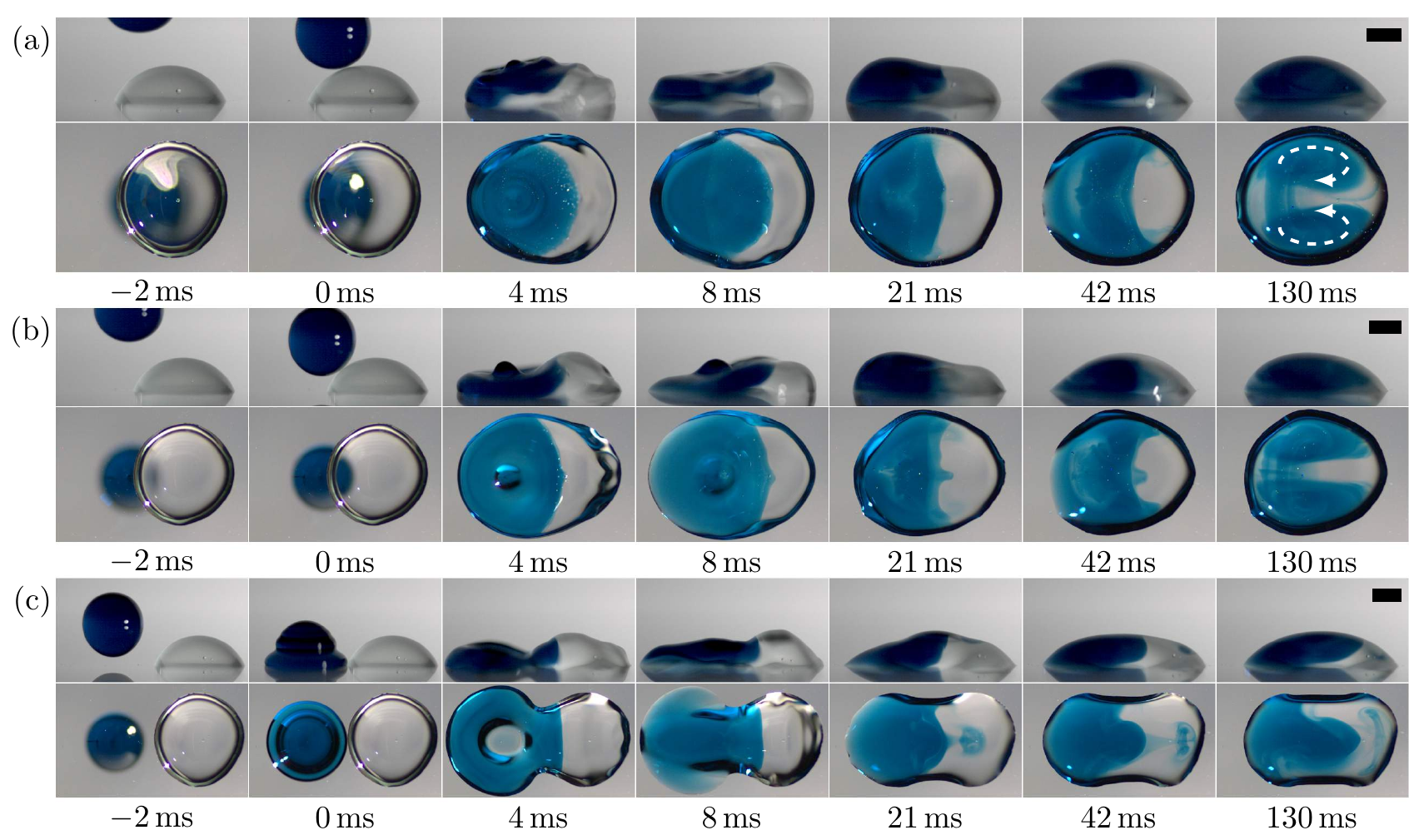}
	\caption{Side and bottom views of a dyed droplet impacting an undyed sessile droplet of the same fluid (fluid 2, 4\% ethanol), for three lateral separations. In panels \panela~and \panelb, the impacting droplet collides with the sessile droplet before the substrate. In panel \panelc, coalescence occurs as the impacting droplet spreads across the substrate. All scale bars are \SI{1}{\milli\meter}.\label{fig:lateralSeparation}}
\end{figure}

For the two smallest lateral separations (Figs.~\ref{fig:lateralSeparation}a and \ref{fig:lateralSeparation}b), the inertia of the dyed droplet significantly disturbs the sessile droplet on impact, generating capillary waves which travel in both directions along the free surface. These capillary waves combined with the spreading of the impacting droplet cause the left contact line to move outwards, which dissipates some energy introduced by the impact \cite{Kapur2007}. The right contact line remains pinned, with the capillary waves insufficient to displace it on this substrate. Right contact line motion may also be inhibited by the outward movement of the left contact line, which commences before the leading capillary wave reaches the right contact line, and draws undyed fluid towards it by mass conservation. After the initial spreading, the left contact line also becomes pinned. Combined with the excess of dyed fluid on the left side of the coalesced droplet, the pinned contact lines induce a recirculatory internal flow as indicated on the \SI{130}{\milli\second} bottom view frame of Fig.~\ref{fig:lateralSeparation}a. Due to this internal flow structure, the dyed fluid is primarily located on the outside of the droplet, whereas the undyed fluid is trapped within the center. Note that such internal flow is not observed in the ostensibly similar experiments of Ref.~\cite{Castrejon-Pita2013}, primarily due to higher Ohnesorge number utilized ($\mathrm{Oh} \approx 0.25$) in that work which yields a reduced influence of surface tension and much greater viscous dissipation.

While recirculatory internal flow alters the distribution of dyed and undyed fluid, it simply advects rather than stretching and folding the internal fluid interface; there is minimal advective mixing. Nevertheless, utilizing both views there does appear to be mixing on the left side of the coalesced droplet (especially visible at \SI{42}{\milli\second} in Fig.~\ref{fig:lateralSeparation}a) due to undyed fluid being propelled into a region where dyed fluid originally resided. Since the precursor droplets consist of the same fluid, the only mechanisms of advective mixing on a short time scale result from the inertia of the impacting droplet and the initial Laplace pressure difference between the coalescing droplets. The inertia derived from these effects is largely dissipated (primarily by viscosity) within a few hundred milliseconds of coalescence. Therefore, molecular diffusion must act over a relatively small internal interface to homogenize the coalesced droplet, which is an extremely slow process. For the millimetric droplets considered here, the estimated time scale to homogenize the droplet is on the order of minutes based on Ref.~\cite{Wilson2018}, during which time significant droplet evaporation would be expected. For such long time scales, internal flows generated by evaporation would provide an additional transport mechanism which may improve the mixing rate \cite{Hu2005}. However, it is clear that achieving good advective mixing is crucial to efficiently realizing a homogeneous coalesced droplet on desirably short time scales.

Despite the difference in lateral separation, the internal flow in Figs.~\ref{fig:lateralSeparation}a and \ref{fig:lateralSeparation}b is remarkably similar. There is a small difference at early times, when penetration of dyed fluid develops along the droplet centerline for the larger lateral separation (Fig.~\ref{fig:lateralSeparation}b), visible at \SI{21}{\milli\second}. This flow structure is located close to the substrate as is clear from the \SI{42}{\milli\second} side view, but does not persist at later times when the internal flow becomes dominated by recirculation. In fact, the only enduring difference between these cases is the droplet footprint on the substrate, with the final droplet shape being closer to a spherical cap in the former case, whereas the footprint is elliptical in the latter. As seen, the difference in droplet footprint does not greatly influence the internal dynamics.

If the lateral separation between the precursor droplets is large enough, then the impacting droplet can land on the substrate before spreading into the sessile droplet to induce coalescence, a situation which may arise when depositing lines or otherwise patterning a substrate \cite{Hsiao2016}. Figure~\ref{fig:lateralSeparation}c presents an experiment with such a lateral separation. Compared to Figs.~\ref{fig:lateralSeparation}a and \ref{fig:lateralSeparation}b, the only experimental difference is in the lateral separation, but the internal flow is significantly different with a jet emanating from the dyed fluid region into the undyed fluid of the precursor sessile droplet, visible at \SI{21}{\milli\second}. From the bottom view, there may appear to be good advective mixing within the coalesced droplet, with significant stretching and some folding of the internal fluid interface. However, the side view shows that the jet is confined to the free surface of the sessile droplet, so there is minimal advective mixing. Similarly, the undyed fluid in the center of the coalesced droplet cannot be perceived from the side view in both Figs.~\ref{fig:lateralSeparation}a and \ref{fig:lateralSeparation}b. Therefore, Fig.~\ref{fig:lateralSeparation} emphasises the need for caution when investigating internal dynamics using only a single view, as has previously been emphasized for mixing \cite{Paik2003electrowetting}.

\subsection{Surface jet formation}
\label{sec:jetFormation}

Internal jets and vortex rings are commonly found in surface tension dominated flows, such as recoiling liquid filaments where they provide a mechanism to escape pinch-off \cite{Hoepffner2013}. However, in Fig.~\ref{fig:lateralSeparation}c the jet is confined to the free surface so a sharp fluid interface is maintained in the bulk. Such surface flows could be utilized to encapsulate a sessile droplet by a second droplet, possibly with different fluid properties \cite{Koldeweij2019}, or to modify its interfacial properties. Alternatively, for droplets deposited to form a continuous line, a sharp transition in line properties may be desired where the presence of such a surface flow could be detrimental. It is therefore of interest to understand the formation of the surface jet in Fig.~\ref{fig:lateralSeparation}c, and whether it can be enhanced or suppressed by modifying the fluid properties. Here, the impacting droplet spreads into the sessile droplet approximately \SI{1.5}{\milli\second} after landing on the substrate to induce coalescence. At this moment, the impacting droplet still has considerable inertia, though some energy has already been dissipated by the displacement of the left contact line. It also has excess surface energy, having not formed a spherical cap, and bears the advancing contact angle which is larger than the equilibrium contact angle. However, at impact (\SI{0}{\milli\second}), the height of the impacting droplet near the point of coalescence is much less than that of the sessile droplet due to its deformed shape. Rapid expansion of the meniscus bridge between the droplets following coalescence generates capillary waves which travel outwards. These capillary waves disturb the free surface of the undyed fluid (visible in the \SI{4}{\milli\second} side view), but are dominated by the ongoing spreading dynamics in the dyed fluid region.

\begin{figure}
	\includegraphics[width=1.5\colwidth]{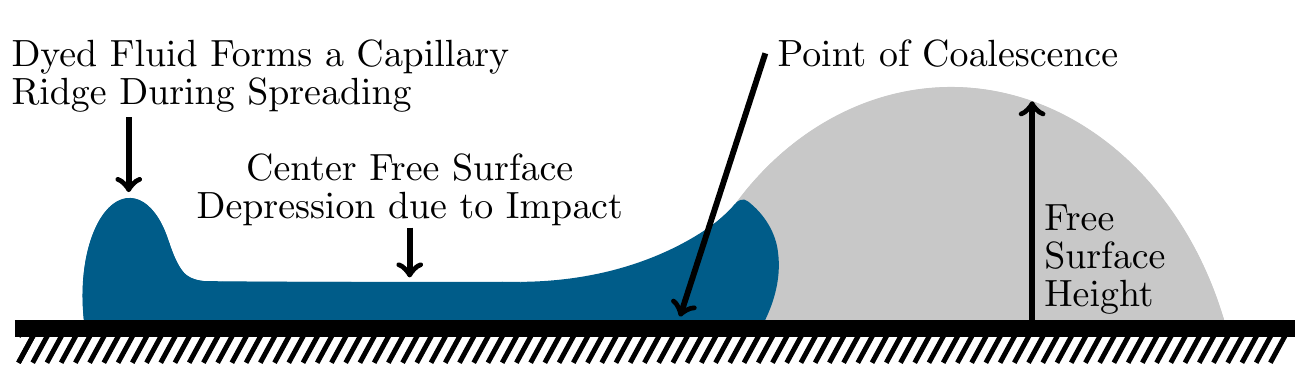}
	\caption{Sketch depicting the coalescence of two droplets of the same fluid at the instance the maximum spread length is reached (typically \SI{3}{\milli\second} after coalescence), represented as a cut-plane through the precursor droplet centers. The impacting droplet lands on the substrate before spreading into the sessile droplet to induce coalescence.\label{fig:maxSpreadCutPlane}}
\end{figure}

Whilst the early coalescence dynamics take place near the meniscus bridge, the dyed fluid continues to spread radially outwards in all other directions until the maximum spread length is reached, which is typically \SI{3}{\milli\second} after coalescence. The spreading dynamics are essentially unaffected by coalescence, except in the immediate region of the point of coalescence. There, the spreading dynamics combine with meniscus bridge growth to push dyed fluid into the region originally occupied by undyed fluid, past the point of coalescence. The impacting droplet otherwise experiences typical deposition dynamics. A large free surface depression develops around its center with a diameter comparable to that of the droplet immediately before impact, while fluid migrates radially outwards to the advancing edges \cite{Rioboo2001}. The resulting free surface topology at the maximum spread length is illustrated in Fig.~\ref{fig:maxSpreadCutPlane} as a cut-plane through the centers of the precursor droplets. The central depression across the dyed fluid is not conspicuous from an external side view due to the axisymmetry of typical deposition dynamics (i.e. it is hidden by the higher outer free surface), but it can be perceived by the relative pixel intensity within the dyed fluid region in the \SI{4}{\milli\second} bottom view frame of Fig.~\ref{fig:lateralSeparation}c. A capillary ridge forms near the contact line since the relatively high advancing contact angle of the substrate prevents further spreading, whilst the radial flow continues to transport dyed fluid outwards to accumulate behind the contact line \cite{Bonn2009}. The generation of such a capillary ridge is critically dependent on the substrate wettability, as on a perfectly wetting substrate the droplet would spread to coat the substrate with a uniform thickness.

\begin{figure}[b]
	\includegraphics[width=1.5\colwidth]{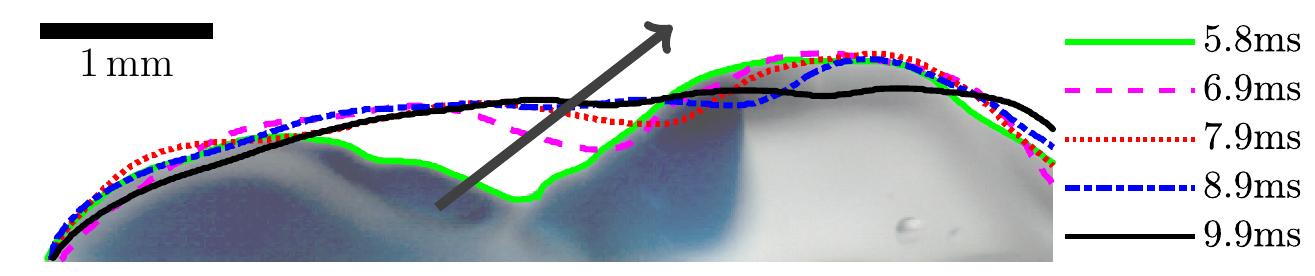}
	\caption{Free surface edges at five early times during the coalescence of two droplets of the same fluid (fluid 2, 4\% ethanol) overlaid onto the \SI{5.8}{\milli\second} side view (faded by 25\%). The gray arrow indicates increasing time. The data correspond to Fig.~\ref{fig:lateralSeparation}c.\label{fig:edgeProgression}}
\end{figure}

The free surface at the maximum spread length (Fig.~\ref{fig:maxSpreadCutPlane}) is severely deformed, whilst the flow is dominated by surface tension. Since the contact line cannot advance further as dyed fluid continues to accumulate in the capillary ridge, fluid is quickly reflected away from the contact line to reduce the free surface area. During the reflection, the left contact angle decreases but the contact line does not recede. Meanwhile, energy is viscously dissipated near the fluid interface due to meniscus bridge growth as dyed fluid is pushed into the undyed fluid region. The latter effect is indicated in Fig.~\ref{fig:maxSpreadCutPlane} by the position of the fluid interface relative to the point of coalescence. Furthermore, the right contact line (of the undyed fluid) remains pinned at this time. The asymmetry in the dynamics resulting from these factors ensures the fluid reflected from the contact line is primarily transported in a single direction towards the undyed fluid along the axis of symmetry between the precursor droplets \cite{Ristenpart2006}. As a result of the left contact line being pinned, the reflected fluid forms a travelling wave rather than simply displacing the contact line to form a spherical cap.

The reflected travelling wave precipitates a progressive increase in free surface height across the depressed free surface of the coalesced droplet. This progression is visible in Fig.~\ref{fig:edgeProgression}, with free surface edges shown at five time instants, overlaid onto the side view at \SI{5.8}{\milli\second}. A large and rapid increase in free surface height precedes the wave, with the free surface height becoming approximately uniform away from the contact lines at \SI{9.9}{\milli\second}. Note that the increase in free surface height occurs wholly within the dyed fluid region of the coalesced droplet. The free surface movement therefore acts to raise dyed fluid within the droplet, which is primarily drawn from the capillary ridge. The upward motion of dyed fluid also draws undyed fluid towards the left contact line closer to the substrate by mass conservation, which generates an overturning internal flow with dyed fluid moving towards the right in the upper part of the droplet. After the travelling wave has passed, the free surface height does not vary significantly, as seen in Fig.~\ref{fig:edgeProgression} where the edges almost overlap close to the left contact line. The wave itself continues across the free surface of the undyed fluid (see the \SI{8.9}{\milli\second} edge in Fig.~\ref{fig:edgeProgression}). However, dyed fluid is not immediately drawn forward with the wave, but the surface jet emanates from the dyed fluid approximately \SI{3}{\milli\second} after the reflected wave has passed over the fluid interface due to a surface flow induced by the preceding dynamics.

\begin{figure}
	\includegraphics[width=1.5\colwidth]{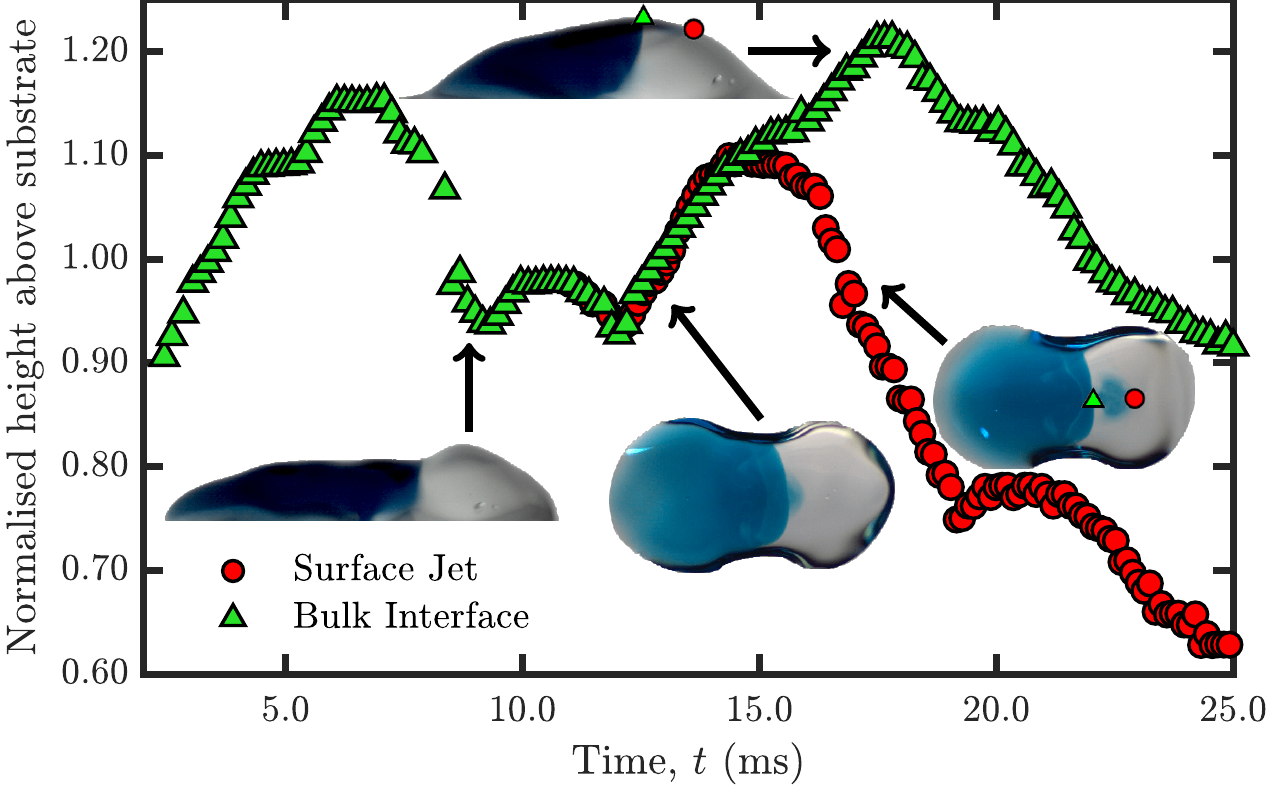}
	\caption{Normalized free surface height for the leading edge positions of the bulk and jet, as indicated on the inset frames. The free surface height is extracted by image processing from each side view frame, matched to the horizontal position determined from the corresponding bottom view frame. Both droplets consist of fluid 2 (4\% ethanol). The data correspond to Fig.~\ref{fig:lateralSeparation}c.\label{fig:verticalPosition}}
\end{figure}

To elucidate the dynamics of surface jet formation, the leading edges of both the bulk fluid interface and surface jet at the free surface were tracked via image processing as described in Sec.~\ref{sec:imageProc}. Figure~\ref{fig:verticalPosition} displays the free surface height corresponding to the horizontal position of these leading edges (see inset frames), normalized by the initial sessile droplet height. The data correspond to Fig.~\ref{fig:lateralSeparation}c, which is a typical example for the prevailing experimental conditions. From the bottom view, only the maximum penetration of dyed fluid is visible, though it is not necessarily uniform across the droplet depth. In particular, the convex nature of the fluid interface depicted in Fig.~\ref{fig:maxSpreadCutPlane} cannot be directly perceived from a bottom view. However, the leading edge of the bulk (at the time the surface jet breaks away) and the surface jet are located close to the free surface which enables them to be accurately tracked.

Figure~\ref{fig:verticalPosition} shows the variation of surface height in time and confirms the free surface at the bulk fluid interface rapidly rises at early times, when the meniscus bridge growth dominates the dynamics. The concurrent spreading dynamics are subordinated to the meniscus bridge growth, though the former acts to push the bulk fluid interface into the originally undyed fluid region, which contributes to the rise in the bulk fluid interface free surface height as the free surface of the undyed fluid is higher than that of the dyed fluid. For the coalescence of symmetric, identical precursor sessile droplets, the free surface height at the fluid interface (directly above the point of coalescence) would be expected to level off and fluctuate around the equilibrium height of the coalesced droplet. However, in Fig.~\ref{fig:verticalPosition} the travelling wave induces a reduction in free surface height at the bulk fluid interface when it approaches approximately \SI{7}{\milli\second} after coalescence. Hence, the travelling wave is characterized by a local depression in the free surface near the bulk fluid interface (visible in Fig.~\ref{fig:verticalPosition}) as the free surface is higher ahead, similar to a breaking wave. Figure~\ref{fig:verticalPosition} therefore shows that the travelling wave passes the bulk fluid interface approximately \SI{9}{\milli\second} after coalescence, after which the free surface at the bulk fluid interface rises and the surface jet forms.

The tracking algorithm automatically identifies the formation of surface structures emanating from the bulk fluid interface (a surface jet here) in the bottom view, then proceeds to track both the bulk and newly formed leading edges simultaneously. As seen in Fig.~\ref{fig:verticalPosition}, the surface jet does not form immediately as the travelling wave passes the bulk fluid interface, nor advances as fast as the travelling wave. These observations indicate that the surface jet forms due to a surface flow induced by the dynamics accompanying the travelling wave. However, the travelling wave not only generates a surface flow, but also an overturning internal flow as noted above. This inference is supported by the convex nature of the bulk fluid interface shortly after the formation of the surface jet, as seen at \SI{21}{\milli\second} in Fig.~\ref{fig:lateralSeparation}c. Indeed, the interface is further right in the upper reaches of the droplet (not just at the free surface), indicating an internal flow in the same direction as the surface jet. The internal flow is quickly damped by viscosity, so the bulk fluid interface becomes stagnant, but the surface flow faces less resistance and endures to generate and transport the surface jet. After the surface jet has formed, the height of its leading edge is initially similar to the bulk fluid interface, but soon decreases (beginning at \SI{15}{\milli\second}) due to the free surface oscillations remaining from the travelling wave, in addition to conventional capillary waves. The corresponding response of the bulk fluid interface in Fig.~\ref{fig:verticalPosition} is delayed relative to the surface jet due to their horizontal separation at this time, with the delay increasing as the surface jet moves further away, but the same trends are observed in both as expected. After \SI{18}{\milli\second}, the height of both tracked leading edges decreases as they progress at different velocities towards the right contact line.

The fluid properties of each precursor droplet in Fig.~\ref{fig:lateralSeparation}c are the same, within experimental error. Hence, the surface jet does not arise due to density differences, which would typically occur at longer time scales and for a larger Bond number \cite{Zhang2015}. There is also no evidence of density-driven stratification even at later times (up to \SI{1}{\second} after coalescence). The surface tensions of the precursor droplets are nominally the same, and hence Marangoni flow is not expected to occur. However, even if the surface tensions were slightly different (i.e. within the experimental error), Marangoni effects do not explain the jet formation. A distinct and well-defined surface jet is observed, which travels exclusively in one direction, rather than spreading to cover the higher surface tension free surface of the undyed fluid which would occur in a Marangoni flow (see Sec.~\ref{sec:surfaceTensionDiff} where a surface tension difference is deliberately introduced). Furthermore, if Marangoni flow were responsible, it would only produce local recirculation in the bulk close to the free surface on the short time scale of surface jet formation, rather than the overturning internal flow observed throughout the depth of the droplet here. Therefore, Marangoni flow can not be the cause of the surface jet seen in Fig.~\ref{fig:lateralSeparation}c (though it can modify or even inhibit the jet, as discussed in Sec.~\ref{sec:surfaceTensionDiff}). These observations substantiate the inference that the surface jet is the result of a surface flow precipitated by a travelling wave reflected from the left contact line.

The primary mechanism which generates the surface jet is the rapid ascent of the depressed free surface (Fig.~\ref{fig:maxSpreadCutPlane}) associated with the impacting droplet, which is enabled by the surface tension dominated flow (low Ohnesorge number). Within the deposition regime, the rate of spreading and the maximum spread length increase with impact velocity. The impact velocity must therefore be sufficient for the droplet to spread far and fast enough that the central free surface depression and capillary ridge can form. A large impact velocity may be detrimental to capillary ridge formation due to the associated increase in the maximum spread length, indicating that an intermediate velocity within the deposition regime is required. The maximum spread length also depends on the advancing contact angle, which was relatively high (approximately \ang{110}) in this work. The substrate wettability is also important after the maximum spread length is reached, as the contact line must remain pinned during fluid reflection to avoid dampening the free surface dynamics and to enable the formation of the travelling wave. Summarizing, the formation of a surface jet depends on the surface tension ratio (a low Ohnesorge number), the impacting droplet velocity (an intermediate Weber number in the deposition regime) and the substrate wettability ($\theta_a \approx \ang{100}$ and pinning here).

\begin{figure}
	\includegraphics[width=1.5\colwidth]{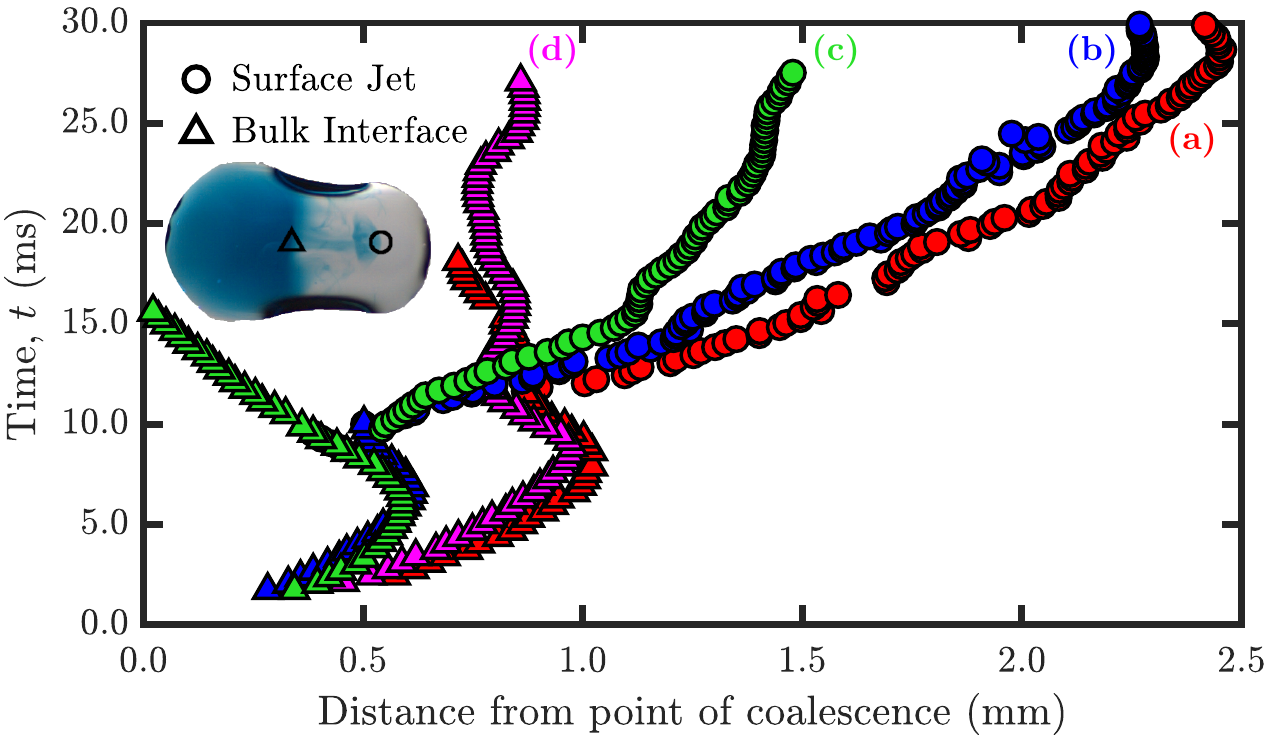}
	\caption{Horizontal position of the bulk and (if applicable) jet leading edges, as indicated on the inset frame from case \panelc, determined from the bottom view frames. Each color (labelled) corresponds to a different experimental case: \panela~Fluid 2 impacts fluid 2 -- Fig.~\ref{fig:lateralSeparation}c. \panelb~Fluid 2 impacts fluid 2 -- Fig.~\ref{fig:changeSurfaceJet}a. \panelc~Fluid 3 impacts fluid 2 -- Fig.~\ref{fig:changeSurfaceJet}b. \paneld~Fluid 2 impacts fluid 3 -- Fig.~\ref{fig:changeSurfaceJet}c.\label{fig:horizontalPosition}}
\end{figure}

\subsection{Surface jet properties}

Figure~\ref{fig:horizontalPosition} shows the horizontal position of the leading edges (both bulk and jet) in time. As seen in Figs.~\ref{fig:lateralSeparation}c and \ref{fig:horizontalPosition}a, the leading edge of the bulk fluid interface initially migrates quickly into the undyed fluid region due to the ongoing spreading dynamics. The interface continues to advance whilst the meniscus bridge grows, but stalls as the travelling wave approaches due to its effect on the free surface. After the travelling wave passes, the leading edge of the bulk fluid interface retracts due to the internal flow identified above. Note that the bulk leading edge is not necessarily on the free surface at this stage, which explains why this retraction can occur despite the advancing internal and surface flow. Figure~\ref{fig:horizontalPosition}a also shows that the jet travels at an almost constant speed across the free surface until it approaches the contact line and is not influenced by fluctuations in free surface height. The bulk fluid interface continues to slowly retract after the surface jet is emitted, with the surface flow continuing to carry dyed fluid in the opposite direction, despite the internal flow in the upper region of the droplet.

The robustness of the surface jet to lateral separation is examined by increasing the lateral separation between the precursor droplets in Fig.~\ref{fig:changeSurfaceJet}a (and the accompanying videos provided in the Supplemental Material \cite{SupplMat}) by \SI{0.32}{\milli\meter} (11\%) compared to Fig.~\ref{fig:lateralSeparation}c, with otherwise identical experimental conditions. Increasing the lateral separation increases the spread length of the impacting droplet at the point of coalescence. However, the spreading dynamics of the impacting droplet are essentially unaffected by coalescence except in the immediate vicinity of the sessile droplet, so the formation of the capillary ridge and the subsequent fluid reflection are not influenced by small changes in lateral separation. Therefore, as the substrate is strongly pinning, any change in the spread length of the coalesced droplet corresponds to a change in lateral separation. Hence, the central depression (Fig.~\ref{fig:maxSpreadCutPlane}) is wider for larger lateral separations and the intrusion of dyed fluid into the originally undyed fluid region due to spreading is less. Nevertheless, a surface jet materializes for both lateral separations, with similar internal and free surface dynamics observed. To elucidate the effect of lateral separation on surface jet propagation, the position of the surface jet leading edges are shown in Figs.~\ref{fig:horizontalPosition}a and \ref{fig:horizontalPosition}b. It can be seen that the increase in lateral separation shifts the position of the bulk fluid interface towards the point of coalescence, and jet formation to an earlier time. However, the propagation of the surface jet is unaffected. Consequently, the formation and propagation of the surface jet is robust to lateral separation in the case that the impacting droplet is deposited on the substrate before spreading into the sessile droplet.

\begin{figure}
	\includegraphics[width=2\colwidth]{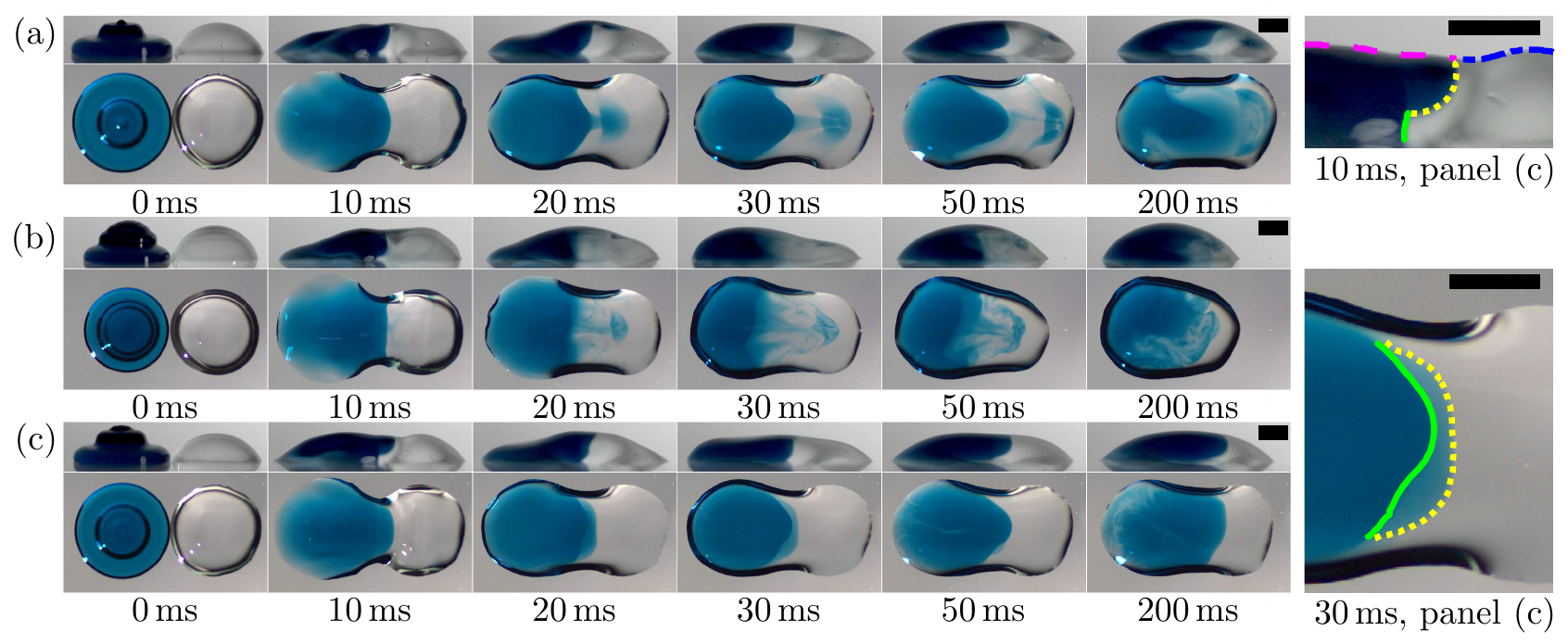}
	\caption{Side and bottom views of a dyed droplet impacting an undyed sessile droplet, with the precursor droplet fluid properties varied between the panels. \panela~A droplet of fluid 2 (4\% ethanol) impacts a sessile droplet of the same fluid. \panelb~A droplet of fluid 3 (8\% ethanol) impacts a sessile droplet of fluid 2 (4\% ethanol). \panelc~A droplet of fluid 2 (4\% ethanol) impacts a sessile droplet of fluid 3 (8\% ethanol). The impacting droplet is always dyed (blue). All scale bars are \SI{1}{\milli\meter}.\label{fig:changeSurfaceJet}}
\end{figure}

\section{Droplets with different surface tensions}

\subsection{Surface flow control}
\label{sec:surfaceTensionDiff}

In this section, a surface tension difference is introduced between the impacting and sessile droplets to initiate a Marangoni flow on coalescence and thereby influence surface jet formation. The impact of a dyed droplet (fluid 3, $\sigma = \SI{50.5}{\milli\newton\per\metre}$) which coalesces with an undyed sessile droplet of higher surface tension (fluid 2, $\sigma = \SI{58.0}{\milli\newton\per\metre}$) is shown in Fig.~\ref{fig:changeSurfaceJet}b and the accompanying videos (provided in the Supplemental Material \cite{SupplMat}). Here, a Marangoni flow arises to reduce the surface area of the undyed fluid which minimizes surface energy. Initially, the Marangoni flow entrains a thin layer of dyed fluid onto the free surface around the outside of the undyed fluid (in the plane of the bottom view), which is visible within \SI{3}{\milli\second} of coalescence. However, the two fluids are miscible, so the small volume of dyed fluid in the film quickly mixes with the undyed fluid below and the surface tension does not change appreciably. The free surface dynamics meanwhile are similar to the equal surface tension case, with the formation of a travelling wave precipitating a rapid rise in the free surface of the coalesced droplet. A surface jet emanates from the dyed fluid region and travels towards the right contact line. However, the induced Marangoni flow also spreads the (higher surface tension) dyed fluid constituting the jet in all directions across the free surface of the undyed fluid. Hence, the Marangoni flow dissipates the inertia of the surface jet, which causes it to stall before reaching the right contact line. The interruption to jet propagation is clear in Fig.~\ref{fig:horizontalPosition}c, where the maximum penetration of the jet is much less than the corresponding case of droplets of the same fluid properties (Fig.~\ref{fig:horizontalPosition}b). The initial speed of the jet is similar though, before it abruptly slows and stalls. With the increased volume of dyed fluid being transported along the free surface due to Marangoni flow, Fig.~\ref{fig:horizontalPosition}c shows that the bulk fluid interface rapidly retracts by mass conservation, in addition to the internal flow identified in the constant fluid properties case. This surface flow induces an internal flow which causes the right contact line to retract whilst the left contact line remains pinned. Marangoni flow also generates additional mixing near to the free surface. Thus the jet penetrates deeper into the coalesced droplet, as visible at \SI{50}{\milli\second} in the side view, with the head of the jet forming a toroidal section. Increased mixing on a short time scale is therefore observed due to the surface tension difference.

To investigate the influence of deposition order, the fluids are swapped between the precursor droplets in Fig.~\ref{fig:changeSurfaceJet}c compared to Fig.~\ref{fig:changeSurfaceJet}b, though the dye remains in the impacting droplet. Hence, the sessile droplet has a lower surface tension (fluid 3, $\sigma = \SI{50.5}{\milli\newton\per\metre}$) than the impacting droplet (fluid 2, $\sigma = \SI{58.0}{\milli\newton\per\metre}$). Marangoni flow therefore opposes the surface flow which typically generates the surface jet, as the formation of a surface jet would increase the overall surface energy. However, the external dynamics are consistent with those in both Figs.~\ref{fig:changeSurfaceJet}a and \ref{fig:changeSurfaceJet}b. Furthermore, the overturning internal flow still arises, which leads to a deformed bulk fluid interface as seen from the image-processed edges in Fig.~\ref{fig:changeSurfaceJet}. The solid green edges indicate where the internal flow is directed towards the left contact line, whereas the dotted yellow edges indicate the depths at which the internal flow (generated by the travelling wave) is towards the right contact line. The internal dynamics are therefore such that a surface jet could form, but not as a result of the opposing Marangoni flow. The suppressed surface flow leaves a distinct, well-defined bulk fluid interface which oscillates around a given horizontal position (Fig.~\ref{fig:horizontalPosition}d). This result demonstrates the influence of deposition order on the internal dynamics when the precursor droplets have different fluid properties and supports the physical arguments surrounding the internal and surface flows made above. It may also be a mechanism underpinning reduced color bleeding previously observed between inkjet printed droplets of different surface tension \cite{Oyanagi2003}.

These results can be elucidated by considering the relative time scales of the inertial and Marangoni flows. Due to the low viscosity and high surface tension of the droplets, coalescence proceeds in the inertial regime after the earliest (sub-microsecond) stage of coalescence \cite{Zhang2015}. The inertial time associated with surface tension driven flow is
\begin{equation}
	\tau_\sigma = \sqrt{\frac{\rho r^3}{\sigma}},
\end{equation}
where $\rho$, $r$ and $\sigma$ are droplet density, radius and surface tension, respectively \cite{Thoroddsen2000}. For a typical droplet in this work (defined in Sec.~\ref{sec:procedure}), $\tau_\sigma \approx \SI{4.5}{\milli\second}$. Note that the inertial time scale relates to the growth of the meniscus bridge between the precursor droplets rather than the dynamics induced by impact and spreading, but it nevertheless provides an indication of the typical inertial time scale and is consistent with the experiments reported. The time scale associated with Marangoni flow is
\begin{equation}
	\label{eq:marangoniTime}
	\tau_m = \frac{\bigl(\mu_o + \mu \bigr)r}{\Delta\sigma},
\end{equation}
where $\mu_o$ is the viscosity of the surrounding air \cite{Kovalchuk2019}. For Figs.~\ref{fig:changeSurfaceJet}b and \ref{fig:changeSurfaceJet}c, $\Delta\sigma \approx \SI{8}{\milli\newton\per\metre}$ and $\mu_o \approx \SI{e-5}{\pascal\second}$ so the corresponding Marangoni time scale is $\tau_m \approx \SI{0.1}{\milli\second}$. That is, the Marangoni time scale is at least one order of magnitude shorter than the inertial time scale for this surface tension difference, which indicates that the action of Marangoni flow is faster and can prevent the formation of the surface jet in Fig.~\ref{fig:changeSurfaceJet}c. 

Note that the inertial and Marangoni time scales are similar ($\tau_\sigma \approx \tau_m$) if
\begin{equation}
	\label{eq:similar}
	\Delta\sigma \approx \bigl( \mu + \mu_o \bigr) \sqrt{\frac{\sigma}{\rho r}} \approx \SI{0.3}{\milli\newton\per\meter}.
\end{equation}
Therefore, Marangoni flow can become important in acting as fast as surface tension generated inertial flows for remarkably small surface tension differences. However, for such small surface tension differences the flow induced may not be strong enough to influence the dynamics despite being able to act quickly, especially if there is another influence on the flow such as the travelling wave in this work. Equation~\ref{eq:similar} nevertheless demonstrates the potential for small surface tension differences to influence internal flows, which could be utilized in the design of devices where larger changes to fluid properties may be undesirable, such as open-surface microfluidics.

\subsection{Regime map}

\begin{figure}
	\includegraphics[width=1.5\colwidth]{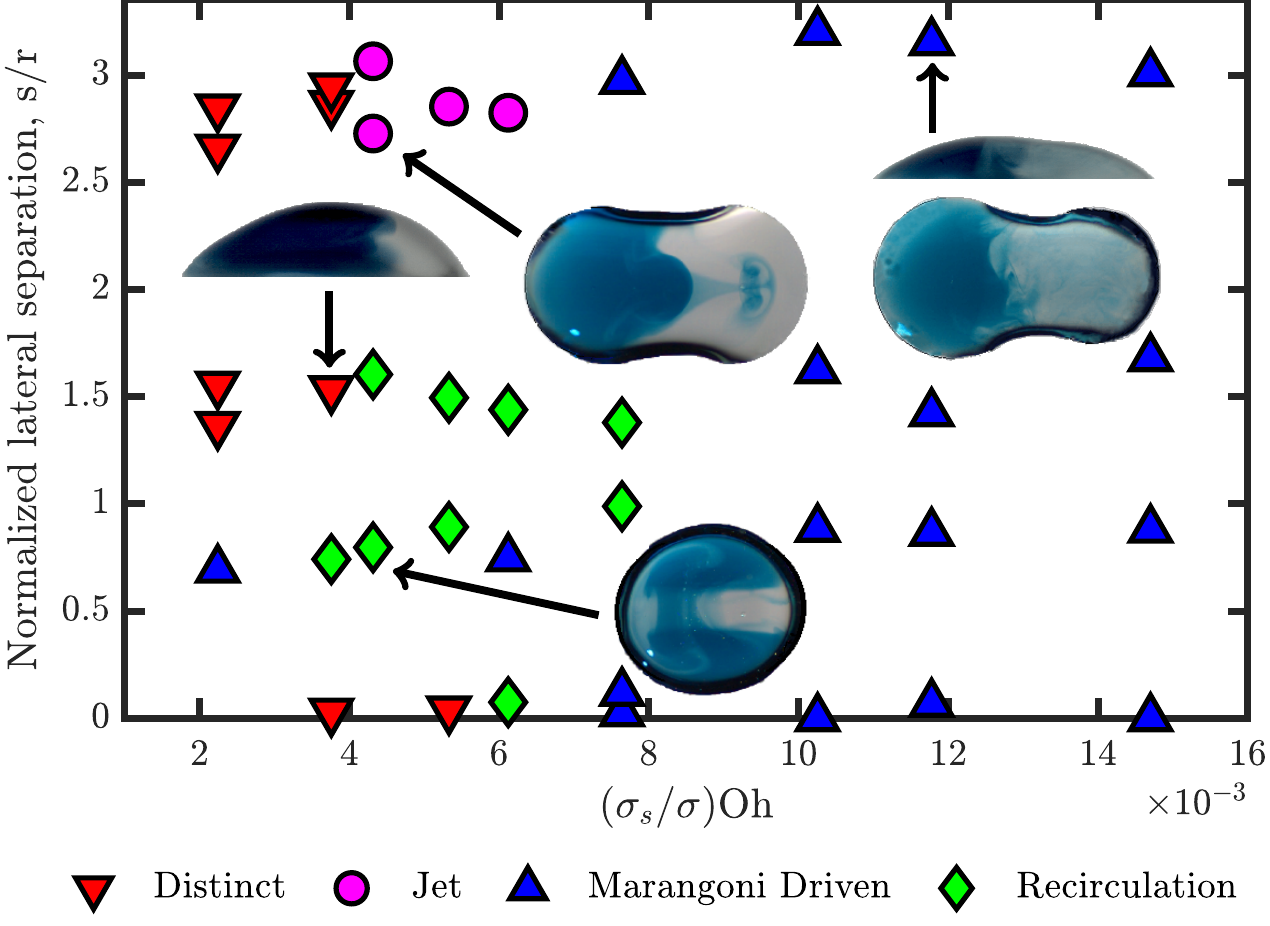}
	\caption{Regime map for the early time flow structures at various lateral separations and different relative droplet fluid properties, characterized by a dimensionless group involving the Ohnesorge number (of the impacting droplet) and the surface tension ratio.\label{fig:regimeMap}}
\end{figure}

To elucidate the conditions in which the previously discussed flow structures arise, Fig.~\ref{fig:regimeMap} presents a regime map which displays the early time flow structures observed at various lateral separations between the precursor droplets, $s$ normalized by the impacting droplet radius, $r$. Denoting the sessile droplet surface tension as $\sigma_s$, the formation of a surface jet depends on the surface tension ratio $\sigma_s/\sigma$ and Ohnesorge number (based upon the impacting droplet properties, so $\mathrm{Oh} \propto \sigma^{-1/2}$) in Sec.~\ref{sec:jetFormation}. Hence, the fluid properties can be characterized by the modified Ohnesorge number $(\sigma_s/\sigma)\mathrm{Oh} \propto \sigma_s\sigma^{-3/2}$ which accounts for both the dominance of surface tension and its difference between the precursor droplets. Each plotted point represents a typical example from at least three repeated experiments of the same case. The qualitative flow description given was consistent between each repeated experiment.

For $\sigma_s \gg \sigma$, vigorous Marangoni flow is quickly induced at all lateral separations, as indicated by Eq.~(\ref{eq:marangoniTime}), preventing larger organized flow structures (e.g. recirculation) from developing. Such cases are described as Marangoni driven and typically result in rapid mixing across the coalesced droplet (discussed in Sec.~\ref{sec:mixing}). Larger flow structures rely on surface tension dominated flow (i.e. low Oh) in addition to a lower surface tension ratio, so typically appear at lower values of $(\sigma_s/\sigma)\mathrm{Oh}$. For $\sigma_s \approx \sigma$, the surface jet appears only at the largest lateral separations (when the impacting droplet hits substrate before the sessile droplet). The flow is dominated by recirculation (see Figs.~\ref{fig:lateralSeparation}a and \ref{fig:lateralSeparation}b) if the lateral separation is smaller for all Ohnesorge numbers studied, as seen in Fig.~\ref{fig:regimeMap} by the clustering of green diamonds. A distinct interface is maintained between the dyed and undyed fluids (e.g. Fig.~\ref{fig:changeSurfaceJet}c) for cases where the sessile droplet surface tension is lower than that of the impacting droplet ($\sigma_s<\sigma$), as explained in Sec.~\ref{sec:surfaceTensionDiff}, shown as red triangles (at low values of the modified Ohnesorge number). Whilst there is rapid mixing driven by a local Marangoni flow in the region of the fluid interface, the interface itself remains sharp due to the suppressed surface flow, without mixing across the whole droplet which occurs in the Marangoni driven cases. A distinct interface can also materialize without surface tension differences for axisymmetric droplet-on-droplet impact ($s/r = 0$) as seen in Fig.~\ref{fig:regimeMap}.

\subsection{Long term dynamics and mixing}
\label{sec:mixing}

The flows considered so far occur on a short time scale. For example, in Fig.~\ref{fig:lateralSeparation}c the surface jet reaches the right contact line less than \SI{30}{\milli\second} after coalescence. Such short term dynamics determine the initial distribution of fluid from each precursor droplet and thus define the initial condition for the longer time scale dynamics which ultimately homogenize the coalesced droplet. Figure~\ref{fig:varySurfaceTensionDiff} and the accompanying videos (see Supplemental Material \cite{SupplMat}) present the coalescence of a dyed, impacting droplet with an undyed, sessile droplet of various relative fluid properties to elucidate the effect of surface tension gradients on the long term dynamics and mixing efficiency. Only the fluid properties of the droplets are varied between each panel in Fig.~\ref{fig:varySurfaceTensionDiff}.

\begin{figure}
	\includegraphics[width=2\colwidth]{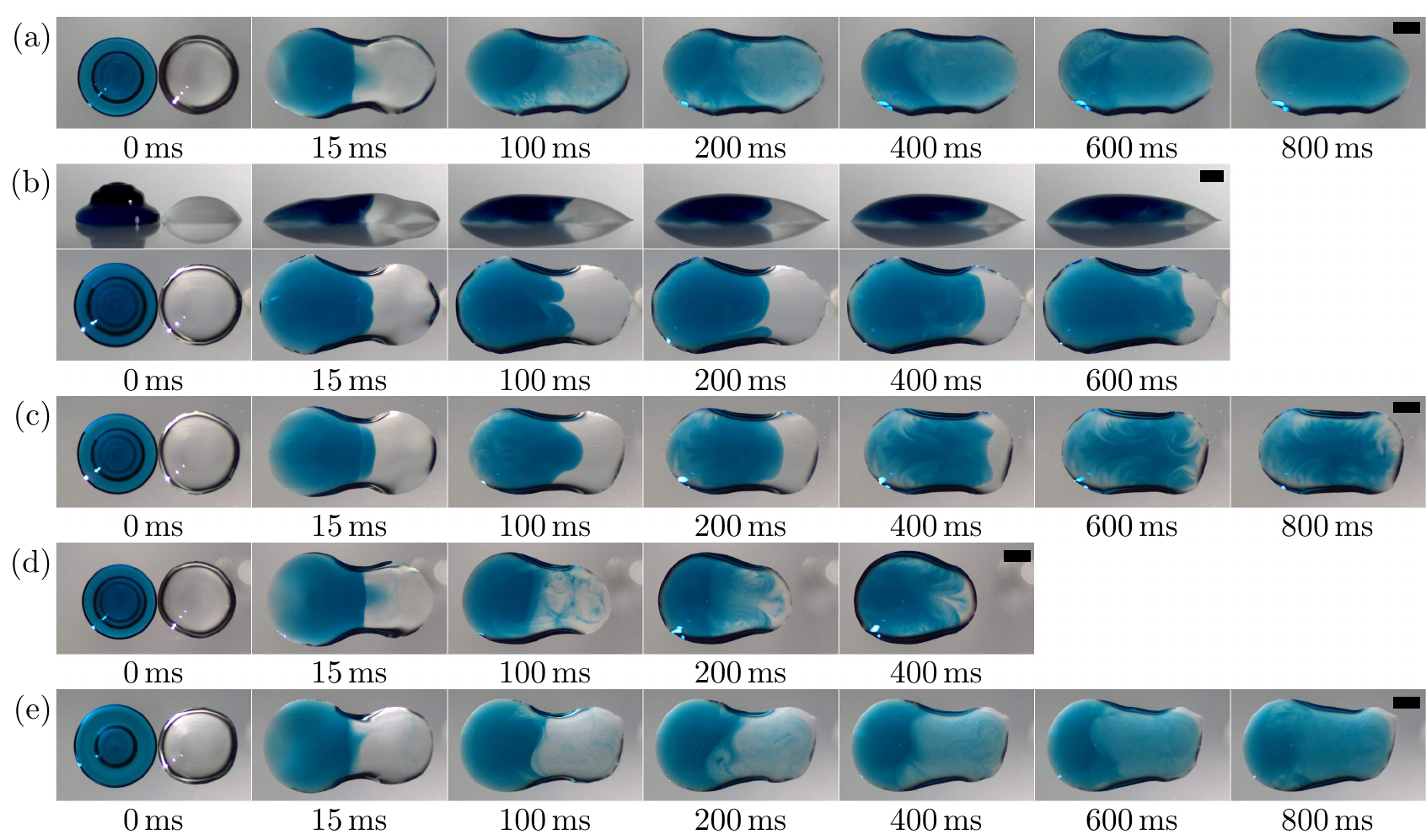}
	\caption{Bottom views of a dyed droplet impacting an undyed sessile droplet, with the fluid properties varied between the panels. The side view is also shown in panel \panelb. \panela~A droplet of fluid 3 (8\% ethanol) impacts a sessile droplet of fluid 1 (water). \panelb~A droplet of fluid 1 (water) impacts a sessile droplet of fluid 3 (8\% ethanol). \panelc~A droplet of fluid 2 (4\% ethanol) impacts a sessile droplet of fluid 3 (8\% ethanol). \paneld~A droplet of fluid 3 (8\% ethanol) impacts a sessile droplet of the same fluid. \panele~A droplet of fluid 4 (18\% ethanol) impacts a sessile droplet of fluid 3 (8\% ethanol). All scale bars are \SI{1}{\milli\meter}.\label{fig:varySurfaceTensionDiff}}
\end{figure}

In Fig.~\ref{fig:varySurfaceTensionDiff}a, the impacting droplet (fluid 3, $\sigma = \SI{50.5}{\milli\newton\per\metre}$) has a lower surface tension than the sessile droplet (fluid 1, $\sigma = \SI{72.4}{\milli\newton\per\metre}$). A surface flow is visible at \SI{15}{\milli\second}, but the large surface tension difference causes dyed fluid to spread over the sessile droplet which arrests it and prevents the formation of a well-defined surface jet. After \SI{100}{\milli\second}, the coalesced droplet is comprehensively covered by dyed fluid with significant mixing near the free surface. The bulk is however not fully mixed as indicated by the non-uniform hue across the droplet in the bottom view. After \SI{800}{\milli\second}, the coalesced droplet appears almost homogeneous and is well mixed. For micrometric droplets ($r\approx\SI{25}{\micro\meter}$) of common fluid properties, complete mixing by diffusion alone is expected after a similar time \cite{Wilson2018}. For the millimetric droplets shown in Fig.~\ref{fig:varySurfaceTensionDiff}a, the surface tension gradient drives vigorous internal flow which improves the efficiency of diffusion to homogenize the coalesced droplet. The fluids are swapped between the precursor droplets in Fig.~\ref{fig:varySurfaceTensionDiff}b compared to Fig.~\ref{fig:varySurfaceTensionDiff}a, with the dye remaining in the impacting droplet. The surface tension gradient suppresses the surface flow, but the overturning internal flow characterized by the deformed bulk interface appears (see also Fig.~\ref{fig:changeSurfaceJet}). The internal fluid interface remains sharp, but rapid mixing (due to the surface tension gradient) causes it to advance quickly through the droplet over the \SI{600}{\milli\second} shown. However, the extent of undyed fluid infiltration into the dyed fluid region is unclear. Compared to Fig.~\ref{fig:varySurfaceTensionDiff}a, there is significantly less mixing after \SI{600}{\milli\second}, which demonstrates that the order of deposition influences the long term dynamics when the precursor droplets have different fluid properties. In particular, the short term dynamics have a considerable influence on the long term mixing efficiency.

The surface tension of the impacting droplet is progressively decreased through the remaining panels of Fig.~\ref{fig:varySurfaceTensionDiff}, with a consistent sessile droplet of fluid 3 ($\sigma = \SI{50.5}{\milli\newton\per\metre}$). In Fig.~\ref{fig:varySurfaceTensionDiff}c, the impacting droplet consists of fluid 2 ($\sigma = \SI{58.0}{\milli\newton\per\metre}$), as in Fig.~\ref{fig:changeSurfaceJet}c. The dynamics are similar to Fig.~\ref{fig:varySurfaceTensionDiff}b, but there is a reduced surface tension difference so Marangoni flow is less prominent which results in slower mixing around the fluid interface. There is also evidence of patterning in the dyed fluid at longer times as undyed fluid moves towards the left contact line, which is not apparent for larger surface tension differences where Marangoni flow homogenizes the fluid in these regions rapidly. In Fig.~\ref{fig:varySurfaceTensionDiff}d, the impacting droplet has the same fluid properties as the sessile droplet (fluid 3, $\sigma = \SI{50.5}{\milli\newton\per\metre}$). A weak surface jet forms which reaches the right contact line, but the surface flow also transports additional dyed fluid across the undyed fluid free surface, as seen at \SI{100}{\milli\second}. Note that this transport of fluid is not spreading due to a Marangoni flow, for which a more uniform film would be expected as seen in Figs.~\ref{fig:varySurfaceTensionDiff}a and \ref{fig:varySurfaceTensionDiff}e. Instead, the central surface flow observed in other image sequences (e.g. Fig.~\ref{fig:changeSurfaceJet}a) becomes wider and less distinct due to the lower surface tension, which transports dyed fluid across a greater proportion of the undyed fluid's free surface. Nevertheless, the distribution of dyed fluid after \SI{200}{\milli\second} indicates recirculation of fluid in a jet-like manner on the free surface, with associated retraction of the right contact line. This result shows that the surface jet becomes narrower and stronger as surface tension increases. While dyed fluid is visible throughout most of the droplet at \SI{400}{\milli\second}, it mostly resides near the free surface in the originally undyed fluid region with relatively little fluid mixing materializing.

In Fig.~\ref{fig:varySurfaceTensionDiff}e, the impacting droplet has a lower surface tension (fluid 4, \SI{39.9}{\milli\newton\per\metre}) than the sessile droplet. A thin film of dyed fluid spreads across the free surface of the undyed fluid due to Marangoni flow, visible at \SI{15}{\milli\second}, but the surface flow generated by impact is not sufficient to transport dyed fluid a significant distance across the free surface or form a surface jet. Compared to Fig.~\ref{fig:varySurfaceTensionDiff}a, the flow is less surface tension dominated which reduces the strength of the surface flow generated by impact. Therefore Marangoni-driven spreading becomes more important and dyed fluid is spread rather than propelled across the free surface. The efficiency of mixing in the coalesced droplet is also reduced due to the lower surface tension, reducing the velocity of the Marangoni-induced internal flow and resulting in the droplet being only partially mixed after \SI{800}{\milli\second}.

These results demonstrate that the relative surface tension between precursor droplets influences the long term dynamics and extent of fluid mixing, in addition to the short term dynamics. Mixing efficiency tends to be greatest when the impacting droplet has a lower surface tension than the sessile droplet, since Marangoni flow augments the surface flow initiated by impact and increases the efficiency of diffusion by extending the internal fluid interface. Comparing Figs.~\ref{fig:varySurfaceTensionDiff}a and \ref{fig:varySurfaceTensionDiff}e, the mixing efficiency increases and the surface flow becomes stronger as the flow becomes more surface tension dominated. The final droplet footprint is also influenced by the relative precursor droplet fluid properties, which may be important in applications requiring precise droplet placement.

\section{Conclusions}

This work has explored in detail flows generated within impacting and coalescing droplets of equal and distinct surface tension, with various lateral separations between the precursor droplets. The fluids used have a high surface tension and low viscosity, leading to surface tension dominated flows exhibiting intricate internal and interfacial dynamics. For precursor droplets of the same fluid properties with small lateral separations, the internal flow within the coalesced droplet is dominated by bulk recirculation due to the impact. However, increasing the lateral separation, such that the impacting droplet first contacts the substrate then spreads into the sessile droplet, results in more complicated internal dynamics and can generate a well-defined surface jet. The surface jet is a robust, repeatable phenomenon that is caused by a reflected wave from the contact line and the capillary ridge that develops there for sufficiently large advancing contact angles. This travelling wave produces internal and surface flow, transporting fluid from the impacting droplet towards the sessile droplet. While the internal flow is rapidly damped by viscosity, the lower resistance at the free surface allows the flow there to continue and generate a surface jet that travels at roughly constant speed towards the opposite side of the coalesced droplet.

The unequivocal identification of the surface jet was only possible by the combination of side and bottom views, since the bottom view only reveals the presence of a jet but not its depth within the droplet. This observation illustrates the need for caution when assessing internal flows and advective mixing from only one view. While confocal microscopy has successfully resolved internal flows and advective mixing at different depths in far more quiescent cases (e.g. Ref.~\cite{Lai2010}), the time scales of the surface tension dominated flows considered in this work are too short to support its use currently.

By modifying the surface tension difference between coalescing droplets, this work shows that surface jets can either be enhanced or suppressed depending on the direction of the resulting Marangoni flow, supported by the derived inertial and Marangoni time scales. Several early-time flow structures are seen, including a sustained distinct separation of the fluid originating in the precursor droplets, or surface jet formation when the surface tension difference is small. For larger surface tension differences, Marangoni flow results in vigorous internal flow which drives different fluids together within the coalesced droplet and contributes to efficient mixing. The conditions for the different flow structures are identified in a regime map expressed in terms of a normalized lateral droplet separation and a modified Ohnesorge number representing the relative droplet fluid properties.

Since the early dynamics determine the distribution of fluid from which longer term mixing dynamics evolve, the order of deposition for droplets of different surface tension is critical for determining the ensuing internal flows and extent of fluid mixing in passively mixed systems. Depositing the higher surface tension droplet first so that the droplet inertia is not opposed by Marangoni flow generally improves mixing efficiency. The final droplet footprint on the substrate can also be affected by the deposition order. These results indicate clear practical implications for printing applications where fluid mixing within droplets is either required or undesired.

\begin{acknowledgments}
	T.C.S. thanks the Department of Engineering Science at the University of Oxford for hosting him during his research visit. The research reported is part of the Leeds Institute for Fluid Dynamics (LIFD), which enabled collaboration between the authors. This work was generously supported by the Engineering and Physical Sciences Research Council (EPSRC) Centre for Doctoral Training in Fluid Dynamics at the University of Leeds [Grant No. EP/L01615X/1], by EPSRC-CBET [Grant No. EP/S029966/1], a Royal Society University Research Fellowship [Grant No. URF\textbackslash{}R\textbackslash{}180016] \& Enhancement Award [Grant No. RGF\textbackslash{}EA\textbackslash{}181002] and a Short Research Visit grant from the EPSRC-funded UK Fluids Network.
\end{acknowledgments}

\bibliography{main}

\end{document}